\documentclass[11pt,twoside]{article}
\usepackage{asp2010}

\resetcounters

\begin{document}

\title{Iris: The VAO SED Application}
\author{S.~Doe,$^1$ N.~Bonaventura,$^1$ I.~Busko,$^2$ R.~D'Abrusco,$^1$ 
	 M.~Cresitello-Dittmar,$^1$ R.~Ebert,$^3$ J.~Evans,$^1$ 
	 O.~Laurino,$^1$ J.~McDowell,$^1$ O.~Pevunova,$^3$ and B.~Refsdal$^1$
\affil{$^1$Smithsonian Astrophysical Observatory, 60 Garden Street,
  Cambridge, MA~02138, USA}
\affil{$^2$Space Telescope Science Institute, 3700 San Martin Drive,
  Baltimore, MD~21218, USA}
\affil{$^3$Infrared Processing and Analysis Center, California
Institute of Technology, Pasadena, CA~91125, USA}}

\begin{abstract}
We present Iris, the VAO (Virtual Astronomical Observatory)
application for analyzing SEDs (spectral energy distributions).  Iris
is the result of one of the major science initiatives of the VAO, and
the first version was released in September 2011.  Iris seamlessly
combines key features of several existing astronomical software
applications to streamline and enhance the SED analysis process.  With
Iris, users may read in and display SEDs, select data ranges for
analysis, fit models to SEDs, and calculate confidence limits on
best-fit parameters.  SED data may be uploaded into the application
from IVOA-compliant VOTable and FITS format files, or retrieved
directly from NED (the NASA/IPAC Extragalactic Database).  Data
written in unsupported formats may be converted for upload using
SedImporter, a new application provided with the package. The
components of Iris have been contributed by members of the VAO.
Specview, contributed by STScI (the Space Telescope Science
Institute), provides a GUI for reading, editing, and displaying SEDs,
as well as defining model expressions and setting initial model
parameter values.  Sherpa, contributed by the Chandra project at SAO
(the Smithsonian Astrophysical Observatory), provides a library of
models, fit statistics, and optimization methods for analyzing SEDs;
the underlying I/O library, SEDLib, is a VAO product written by SAO to
current IVOA (International Virtual Observatory Alliance) data model
standards.  NED is a service provided by IPAC (the Infrared Processing
and Analysis Center) at Caltech for easy location of data for a given
extragalactic astronomical source, including SEDs.  SedImporter is a
new tool for converting non-standard SED data files into a format
supported by Iris.  We demonstrate the use of SedImporter to retrieve
SEDs from a variety of sources--from the NED SED service, from the
user's own data, and from other VO applications using SAMP (Simple
Application Messaging Protocol).  We also demonstrate the use of Iris
to read, display, select ranges from, and fit models to SEDs.
Finally, we discuss the architecture of Iris, and the use of IVOA
standards so that Specview, Sherpa, SEDLib and SedImporter work
together seamlessly.

\end{abstract}

\section{Introduction}
The VAO (Virtual Astronomical
Observatory)\footnote{\url{http://www.usvao.org/}} is the US VO
(Virtual Observatory) effort, funded by the NSF and NASA.  In its
first year of development, the VAO has worked on several projects to
support its major science initiatives \citep{O18_adassxxi}.

Our teams at SAO, STScI and IPAC have collaborated to support analysis
of SEDs (spectral energy distributions).  The results are
Iris,\footnote{\url{http://cxc.cfa.harvard.edu/iris/}} an application
for SED analysis; SedImporter, a tool to make SEDs from data collected
from many sources; a SED service\footnote{\url{http://vo.ned.ipac.caltech.edu/services/accessSED?REQUEST=getData&TARGETNAME=:targetName}} provided by
NED\footnote{\url{http://ned.ipac.caltech.edu/}}; and a library,
SEDLib,\footnote{\url{http://cxc.cfa.harvard.edu/iris/sedlib/}}
providing functions to read and write VO-compliant files.  The VAO
distributes Iris and SedImporter in a single package, for Linux and
Mac OS X systems.

In this paper, we discuss importing SEDs with SedImporter; retrieving
SEDs from NED; and analyzing SEDs in Iris.  We also discuss the IVOA
standards and protocols that allowed us to provide this software and
service, using code from previous projects.

\section{Build a SED with SedImporter}
The new SedImporter tool provides functions to build SEDs.
SedImporter can read and write SEDs in formats compliant with the IVOA
Spectrum Data Model,
v1.03.\footnote{\url{http://ivoa.net/Documents/cover/SpectrumDM-20071029.html}}

\begin{figure}[!ht]
\center{\includegraphics[scale=0.3]{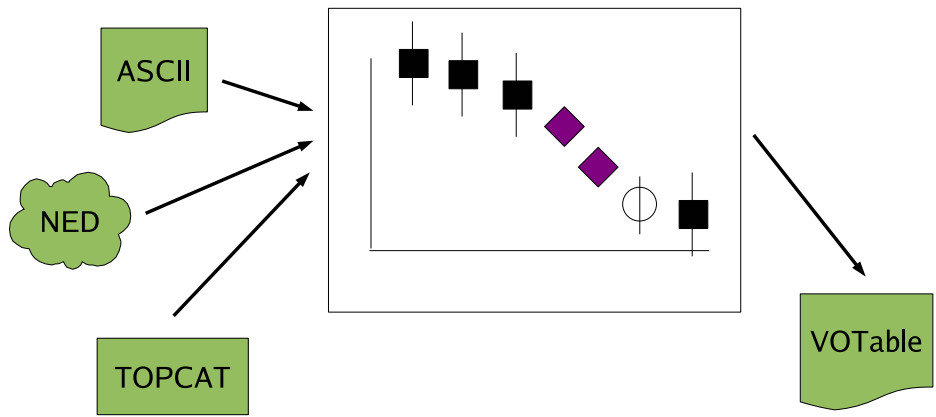}}
\caption{SedImporter combines SED data from different sources.}
\label{fig:F3_1}
\end{figure}

SedImporter also provides functions to read data from non-compliant
files; and also has a
SAMP\footnote{\url{http://ivoa.net/Documents/SAMP/20101216/index.html}}
interface, to import data from other VO-enabled applications (see
Figure~\ref{fig:F3_1}).  SedImporter can write the SED to a file, or
send it to Iris via SAMP.

\section{Analyze a SED with Iris}
Iris provides functions for displaying SEDs and their metadata, and
selecting data ranges for analysis.  It also provides a set of basic
models, which can be combined to model continuum and features.  Iris
can fit models to data and report back fit results.

\begin{figure}[!ht]
\center{\includegraphics[scale=0.31]{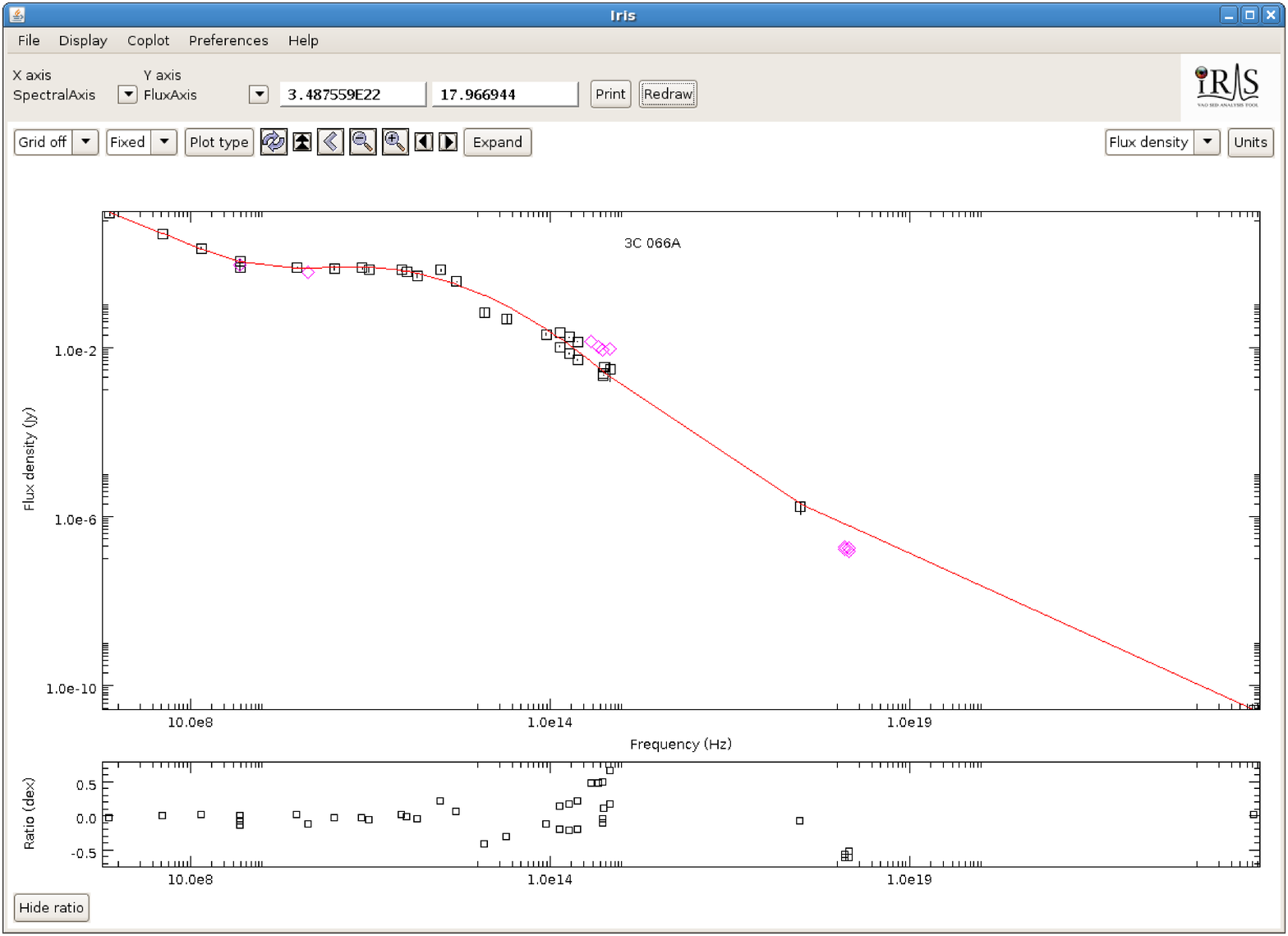}}
\caption{Iris fit to the SED for 3C 066A.  The best-fit model is indicated in
  red.}
\label{fig:F3_2}
\end{figure}

Figure~\ref{fig:F3_2} shows the best fit for 3C 066A, for the sum of a
broken power-law and a logparabola model.  The best-fit model can be
saved and used in future Iris sessions.

\section{Structure of Iris and SedImporter}
\begin{figure}[!ht]
\center{\includegraphics[scale=0.3]{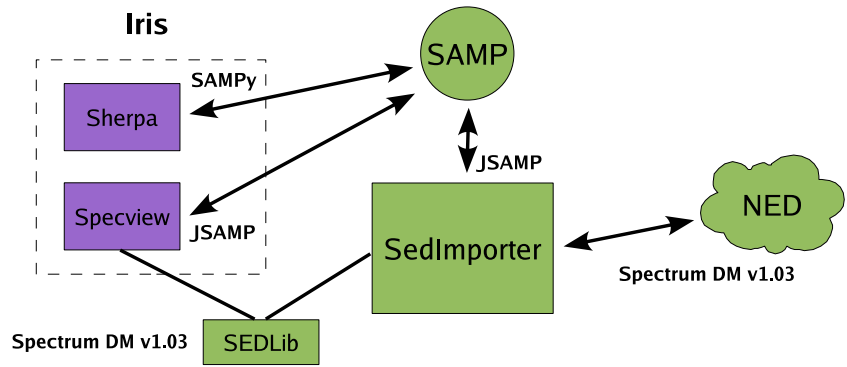}}
\caption{Structure of Iris and SedImporter.}
\label{fig:F3_3}
\end{figure}

Iris and SedImporter include code from mature projects,
bound with code implementing IVOA standards and protocols.
Figure~\ref{fig:F3_3} shows a high-level view of their structure.

SEDLib is a Java library, providing functions to read and write SED
files, in compliance with the Spectrum DM v1.03.  Both SedImporter and
Iris link to this library.

Iris is composed of two applications: Specview (GUI) and Sherpa
(fitting).  Communication is managed via SAMP (with Sherpa using the
SAMPy module,\footnote{\url{http://pypi.python.org/pypi/sampy/}} and
Specview and SedImporter using
JSAMP\footnote{\url{http://software.astrogrid.org/doc/p/samp/1.3/}}).
To the user, Iris acts like a single application.

NED provides a SED service for the objects in its database.  Given a
target name or sky coordinates, NED returns a SED file that complies
with Spectrum DM v1.03.

\section{Conclusions}
In our first year, our SED team has produced a NED SED service, a tool
for building VO-compliant SEDs, a library for developers, and an
application for analyzing SEDs.  We have done this using existing
software coupled with IVOA standards and protocols.

In our second year of development we plan to add to Iris
science functionality, refine its GUI, and comply with updates to the
IVOA Spectrum and SED Data Models.

\acknowledgements Support for the development of Iris/SedImporter is
provided by the Virtual Astronomical Observatory contract AST0834235.
Support for Sherpa is provided by the National Aeronautics and Space
Administration through the Chandra X-ray Center, which is operated by
the Smithsonian Astrophysical Observatory for and on behalf of the
National Aeronautics and Space Administration contract NAS8-03060.
Support for Specview is provided by the Space Telescope Science
Institute, operated by the Association of Universities for Research in
Astronomy, Inc., under National Aeronautics and Space Administration
contract NAS5-26555.  This research has made use of the NASA/IPAC
Extragalactic Database (NED) which is operated by the Jet Propulsion
Laboratory, California Institute of Technology, under contract with
the National Aeronautics and Space Administration.

\bibliography{F3}

\begin{thebibliography}{}
\expandafter\ifx\csname natexlab\endcsname\relax\def\natexlab#1{#1}\fi
\expandafter\ifx\csname url\endcsname\relax
  \def\url#1{\texttt{#1}}\fi
\expandafter\ifx\csname urlprefix\endcsname\relax\def\urlprefix{URL }\fi
\providecommand{\eprint}[2][]{\url{#2}}

\bibitem[{Hanisch et~al.(2012)}]{O18_adassxxi}
Hanisch, R., et~al. 2012, in ADASS XXI, edited by P.~Ballester, \& D.~Egret
  (San Francisco: ASP), vol. TBD of ASP Conf. Ser., TBD

\end{thebibliography}

\end{document}